\documentclass[12pt]{article}

\usepackage{scicite}

\usepackage{float}
\usepackage{times}

\usepackage{epsfig}
\usepackage{graphicx}
\usepackage{amsmath,amssymb}

\newcommand{\beq}[1]{
\begin{equation}
\label{e#1} }

\newcommand{\eeq}{
\end{equation}
}

\newenvironment{sciabstract}{%
\begin{quote} \bf}
{\end{quote}}

\newcounter{lastnote}

\title{Inertial displacement of a domain wall excited by ultra-short circularly polarized laser pulses}

\author
{T.~Janda,$^{1}$ P.~E.~Roy,$^{2}$  R.~M.~Otxoa,$^{2}$ 
\\Z.~ \v{S}ob\'{a}\v{n},$^{3}$ A.~Ramsay,$^{2}$  A.~C.~Irvine,$^{4}$ F.~Trojanek,$^{1}$ R.~P.~Campion,$^{5}$  B.~L.~Gallagher,$^{5}$ 
\\P.~N\v{e}mec,$^{1}$ T.~Jungwirth$^{3,5}$ J.~Wunderlich,$^{3,2,\#}$ \\
\\
\normalsize{$^{1}$Faculty of Mathematics and Physics, Charles University,}\\ 
\normalsize{Ke Karlovu 3, 121 16 Prague 2, Czech Republic}\\
\normalsize{$^{2}$Hitachi Cambridge Laboratory, J. J. Thomson Avenue,}\\ 
\normalsize{Cambridge CB3 0HE, United Kingdom}\\
\normalsize{$^{3}$Institute of Physics, Academy of Sciences of the Czech Republic,}\\ 
\normalsize{Cukrovarnick\'a 10, 162 00 Praha 6, Czech Republic}\\
\normalsize{$^{4}$Microelectronics Research Centre, Cavendish Laboratory, University of Cambridge,}\\
\normalsize{Cambridge CB3 0HE, United Kingdom}\\
\normalsize{$^{5}$School of Physics and
Astronomy, University of Nottingham,}\\
\normalsize{Nottingham NG7 2RD, United Kingdom}\\
\\
\normalsize{$^\#$To whom correspondence should be addressed; E-mail:  jw526@cam.ac.uk.}
}

\date{\today}
\begin{document} 


\baselineskip24pt


\maketitle 

\begin{sciabstract}

Domain wall motion driven by ultra-short laser pulses is a prerequisite for envisaged low-power spintronics combining storage of information in magneto-electronic devices with high speed and long distance transmission of information encoded in circularly polarized light.  Here we demonstrate the conversion of the circular polarization of incident femtosecond laser pulses into inertial displacement of a  domain wall in a ferromagnetic semiconductor. 
In our study we combine electrical measurements and magneto-optical imaging of the domain wall displacement with micromagnetic simulations. The optical spin transfer torque acts over a picosecond recombination time of the spin-polarized photo-carriers which only leads to a deformation of the internal domain wall structure. 
We show that subsequent depinning and micro-meter distance displacement without an applied magnetic field or any other external stimuli can only occur due to the inertia of the domain wall. 
\end{sciabstract}



DW motion driven by magnetic field or electrical current have been shown to be promising concepts for the development of new logic \cite{Allwood2005},  sensing \cite{Diegel2007},  and memory \cite{Parkin2008} devices.  Transient effects giving rise to DW inertia are among the intriguing basic science problems in this research area and have important implications for the performance of DW devices \cite{Saitoh2004,Thomas2006,Thomas2010,Rhensius2010,Vogel2012,Stein2013}. In general, magnetization dynamics is described by first-order differential equations. Moving DWs can be, however, modeled and can behave in experiments as effective particles with inertia whose microscopic origin is in the transient deformation of the DW internal structure when excited out of equilibrium by the magnetic field or current.  

DWs driven by short field \cite{Rhensius2010} or current \cite{Thomas2010} pulses of length $\sim1-10$~ns and moving at characteristic velocities reaching  $\sim0.1-1$~$\mu$m/ns \cite{Yang2015a} are displaced over the duration of the pulse by distances at least comparable but typically safely exceeding the domain wall width. In this regime inertia, causing a delayed response to the driving field and a transient displacement after the pulse, is not the necessary prerequisite for the device operation and is rather viewed as negative factor. It can set the operation frequency limit of the DW device and potentially affect precise positioning of the DW by the driving pulse. Realizing massless DW dynamics is therefore one of the goals in the research of field-driven and current-driven DWs \cite{Vogel2012}.

The aim of our study is the demonstration of a micrometer-scale DW displacement by circularly-polarized, ultra-short laser pulses (LPs). Our experiments are in the regime where the external force generated by the LP acts on the picosecond time-scale over which the expected sub-nanometer DW displacement would be orders of magnitude smaller than the DW width and insufficient for any practical DW device implementation. Inertia allowing for a free transient DW motion after the ultra-short optical excitation is the key for this regime of operation of the opto-spintronic DW devices. 

Our study links the physics of inertial DW motion with the field of optical recording of magnetic media. The manipulation of magnetism by circularly polarized light, demonstrated already in ferrimagnets \cite{Stanciu2007},  transition metal ferromagnets \cite{Lambert2014},  and ferromagnetic semiconductors \cite{Ramsay2015},  has become an extensively explored alternative to magnetic field or current induced magnetization switching. Our work demonstrates that optical recording can in principle be feasible at low power when realized via an energy-efficient DW displacement driven by ultra-short LPs and without the need to heat the system close to the Curie temperature. 

The III-V based ferromagnetic semiconductor used in our study is an ideal model system for the proof of concept demonstration, as well as, for the detailed theoretical analysis of the DW dynamics in this new regime. DWs in the out-of-plane magnetized (Ga,Mn)(As,P) have a simple Bloch wall structure with low extrinsic pinning \cite{DeRanieri2013}. The non-thermal optical spin transfer torque (oSTT) mechanism which couples the circular polarization of the incident light to the magnetization via spin-polarized photo-carriers is microscopically well understood in this ferromagnetic semiconductor material \cite{Nemec2012}. In our experiments, individual circularly polarised $\sim100$~fs short LPs at normal incidence and separated by $\sim10$~ns expose an area with a single DW. As illustrated in Fig. 1A, the generated perpendicular-to-plane spin-polarised photoelectrons exert the oSTT only in the region with an in-plane component of the magnetization, i.e. in the DW.  The action of the oSTT is limited by the photoelectron recombination time $\sim10$~ps. 

To probe the inertial DW motion, we make use of elastic properties of a coherently propagating DW. 
First, the Oersted field generated in a stripe line above the magnetic bar nucleates a reversed magnetic domain. Then, a single DW is driven towards a cross structure by a small external magnetic field of a slightly larger magnitude than the propagation field $B_{PR}$. The low $B_{PR}$ of $\sim0.1$~mT found in our bar devices patterned from an epitaxially grown Ga$_{0.94}$Mn$_{0.06}$As$_{0.91}$P$_{0.09}$ 25 ~nm thick film implies a very small DW pinning on structural defects and inhomogeneities. In this case, DW propagation is coherent and a straight DW becomes pinned at the entrance of the cross structure as shown in Fig. 1B.

To continue the DW propagation through the cross, the DW must increase its length which is accompanied by an increase in its magnetic energy. This results in a restoring force which can be expressed in terms of a virtual restoring field $B_R(x)$ that depends on the position $x$ of the DW. Here, $B_R(x)$ acts as to always drive the DW back to the cross entrance. The magnetic field driven expansion of a DW pinned at the cross entrance is analogous to the inflation of a two-dimensional soap-bubble (see Fig. 2A). The DW depins when the applied field exceeds the maximum restoring field $B_R^{max}$ \cite{Wunderlich01}. Within this model, $B_R(x)$ reaches its maximum value $|B_R^{max}|=\sigma/(M_S\cdot w)$ at the cross  center at $x=0$ (Fig 2 B) and the DW can only depin once it passes the cross center.  Here, $\sigma = 4\sqrt{AK_{E}}$  is the DW energy per unit area, $K_{E}$ the effective perpendicular anisotropy coefficient, $A$ the exchange stiffness, $M_S$ the saturation magnetization and  $w$ is the width of the bar.

The DW can be depinned from the cross by either an applied magnetic field $B_A > |B_R^{max}|$ or by the oSTT. We can therefore use $|B_A| \leq |B_R^{max}|$ to calibrate the strength of the oSTT.

First, however, we have to confirm the elastic nature of DWs in our devices, and verify the applicability of the bubble-like DW model of Fig. 2A.
For this we performed magnetic field driven DW motion experiments without optical excitation. Depinning fields for three different devices with bar widths of  2, 4 and 6~$\mu$m are shown in Fig 2C as a function of the inverse bar width. The slope of the linear fit agrees with that obtained from the measured effective perpendicular anisotropy, $K_E = 1200$~Jm$^3$, the saturation magnetization, $M_S =18$~kA/m, and assuming the exchange stiffness, $A = 50$~fJ/m, which is a reasonable estimate for our GaMnAsP film \cite{DeRanieri2013}. 
The elastic behaviour of the  $\pi \sqrt{A/K_{E}} \sim 20$~nm wide DW is also confirmed by  MOKE images of the $6 ~\mu$m  wide bar device shown in panels (i)-(iii) of Fig 2D . In panel (i), the DW bends into a bubble-like shape under the influence of an applied field $B_A$ = $0.25$~mT. Panel (ii) shows that the restoring field drives the DW back to the cross-entrance after $B_A$ is turned-off. Panel (iii) displays the difference between the two MOKE images (i) and (ii), confirming the bubble like shape of the DW. In addition, anomalous Hall effect (AHE) measurements performed on the 4Å~$\mu$m  device under alternating field excitation $B_A (t)=B_0|sin(w t)|$ also confirm the elastic DW behaviour (Fig 2E). If $B_0$ does not exceed $|B_R^{max}|$, e.g., for $B_0 = 0.2$~mT (green), and $B_0 = 0.3$~mT (blue), the periodic variation of the AHE signal indicates that the DW is at the position $x$ where  $B_R(x)$ and $B_A (t)$ compensate. The residual AHE signal at $B_A = 0$ of about 10$~\%$ of the maximum AHE signal at reversed saturation (DW depinned from the cross) corresponds to the AHE-response for the magnetization distribution with a straight DW located at the cross entrance. (For more details see Supplementary information.)

We now combine the elastic pinning properties of the DW at the cross with the light induced excitation experiments in order to proof the inertial character of the oSTT-induced DW motion. The basic idea of our experiment is to exploit the elastic restoring force which is acting {\textit{continuously} throughout the entire $\sim1~\mu$m wide cross against the expansion of the DW which is driven by {\textit{individual} $\sim100$~fs LPs. The photo-generated electrons can transfer their spin to the magnetization only during their $\sim10$~ps lifetime which is 3 orders of magnitude shorter than the pulse separation time of $\sim10$~ns. 

Our experiments are performed at $90~K$ sample temperature.  LPs with  a wavelength $\lambda$ = 750~nm excite photo-electrons slightly above the bottom of the GaAs conduction band so that for a circularly polarized incident light, photo-electrons become spin-polarized with the degree of polarization approaching the maximum theoretical value of $50\%$ \cite{Pikus1984}. 
To avoid the difficulty with aligning our $\sim1~\mu$m Gaussian spot on top of a  $\sim20$~nm wide DW, we employ the experimental procedure sketched in Fig 3.A. First, a straight DW is positioned at the cross entrance. Then, the LP spot is placed $10~\mu$m away from the DW on the reversed domain side. The spot is then swept at a rate of $\sim2~\mu$m/ms  for 20$~\mu$m along the bar so that the initial DW position is crossed by the spot  and approximately $\sim 10'000$ ultra-short LPs time-separated by $\sim$10~ns expose the DW.  

The dependencies of the depinning field $B_{DP}$ on the  LP energy density for circularly polarized $\sigma^{+}$, $\sigma^{-}$ and linearly polarized $\sigma^{0}$ LPs are shown in Fig 3.B. $B_{DP}$ corresponds to the lowest applied magnetic field necessary to depin the DW from the cross and is different from $-B_R^{max}$ due to the LPs. First, we recognize  a reduction of  $B_{DP}$  with increasing energy density for all three LP polarizations. In case of the linear polarization, i.e.,  without oSTT contributions, we attribute the reduction of  $B_{DP}(\sigma^{0})$ only to the LP induced sample heating. For circularly polarized LPs, additional contributions from the oSTT are present.  We observe for all measured LP energy densities that $B_{DP}(\sigma^{+}) < B_{DP}(\sigma^{0}) < B_{DP}(\sigma^{-})$  for the positive magnetization orientation of the nucleated domain. In case of $\sigma^{+}$ polarised LPs and at high enough LP energy densities (above 12~mJ/cm$^2$) the DW depins without an applied magnetic field  (and even at small negative applied magnetic field which opposes DW expansion). For  $\sigma^{-}$ polarized LPs and the same initial domain configuration, we do not observe the zero-field DW depinning up to the highest  LP energy density used in our experiments (see inset of Fig. 3B) \cite{heating_experiment}. 

The differential MOKE image in Fig. 3C shows an example of the domain configuration after the DW has depinned from the cross entrance by optical excitation in conjunction with a constant applied magnetic field $B_{A}$ which is larger than the DW propagation field of the bar outside the cross. After depinning from the cross irradiated by polarized LPs, the DW becomes pinned again at a second cross 
which was not irradiated during the experiment. 
Fig. 3D shows the final domain configuration after DW depinning by $\sigma^{+}$  polarised LPs at zero applied magnetic field. In this case, the 
$\sigma^{+}$ polarized LPs depin and drive the DW forward to the final irradiated spot position. 

From the measurements shown in Fig 3 we can conclude that for the given initial domain configuration, the oSTT generated by $\sigma^{+}$  ($\sigma^{-}$) polarised LPs assists (opposes) DW depinning by driving the DW towards (away) from the cross center where the pinning strength described by $B_R(x)$ becomes maximal. Considering the $\sim100$~fs  short and $\sim$10~ns time-seperated LPs, depinning of the DW by the oSTT becomes only possible if the elastically pinned DW propagates forward in between successive LPs. Depinning by a DW motion without inertia would require DW velocities of more than 1~$\mu$m/ns which are unrealistically high for DW motion in GaMnAsP films \cite{DeRanieri2013}.

To verify our interpretation, we repeated our measurements at the inverted magnetization configuration in which the reversed magnetization of the nucleated domain points in negative ($- m^R_z$) direction. In this case, the oSTT should act in the opposite direction. Indeed, we observe the opposite helicity dependency in our experiments. Fig.4 shows measurements on a $4~\mu $m wide device comparing the two magnetisation configurations. The consistency found between  $B_{DP}(\sigma^{+(-)}, +m^R_z) \approx -B_{DP}(\sigma^{-(+)}, -m^R_z) $ and $B_{DP}(\sigma^{0}, +m^R_z) \approx -B_{DP}(\sigma^{0}, -m^R_z)$ confirms oSTT mechanism and the high reproducibility of our measurements.

Note, that a heat-gradient can in principle also drive the DW motion \cite{Tetienne2014}. The heat-gradient driven motion can become helicity dependent if the light absorption in the two adjacent magnetic domains is helicity-dependent due to the magnetic circular dichroism (MCD). In our experiments, such a scenario is unlikely because about $\sim$ 98\%  of the LP light penetrates through the 25nm thick magnetic GaMnAsP film and is absorbed and transformed into the heat in the GaAs substrate with no dependence on the helicity. In the Supplementary information we present also helicity-dependent DW experiments at photon-energies ranging from below the band-gap up to high energies where the net spin-polarization of photo-electrons is reduced due to the excitation from the spin-orbit split-off band. Since we do not observe the helicity-dependent DW depinning at photon-energies where MCD of GaMnAsP is still present while simultaneously the photoelectron polarisation is strongly reduced, we conclude that MCD is not the origin of the observed helicity dependent DW depinning. 

To further exclude heat-gradient related DW drag effects due to non-uniform heating by the Gaussian-shaped LP spot, we have performed measurements with opposite laser spot sweep directions. In this case, the heat-gradient with respect to the initial DW position is inverted. As shown in the Supplementary information, sweeping the LP spot  along the bar from an initial position outside of the nucleated domain to the final position in the nucleated domain does not change the helicity dependency of the depinning field.
Additional measurements on devices with 2 and 6~$\mu$m wide bars have, apart from the stronger (weaker) DW pinning strength and larger (smaller) temperature increase from LP heating in the 2~$\mu$m (6~$\mu$m) device, also confirmed that $B_{DP}(\sigma^{+(-)}) < B_{DP}(\sigma^{0}) < B_{DP}(\sigma^{-(+)})$ for $+(-)m^R_z$.

We now support our interpretation of the experiment by 1-dimensional Landau-Lifshitz-Bloch (LLB) numerical simulations of the magnetization $\bold{m}$ \cite{Schieback2009}, coupled to the precessional dynamics of the spin-polarized photo-carrier density, $\bold{s}$ \cite{Nemec2012}:
\begin{center}
\begin{eqnarray}
\frac{\partial\bold{m}}{\partial t}=-\gamma\bold{m}\times\bold{H}_{\text{eff}} -\frac{\gamma\alpha_{\perp}}{m^{2}}\bold{m}\times\left(\bold{m}\times\bold{H}_{\text{eff}}\right)+\frac{\gamma\alpha_{||}}{m^{2}}\left(\bold{m}\cdot\bold{H}_{\text{eff}}\right)\bold{m}
\end{eqnarray}
\end{center}

\begin{center}
\begin{eqnarray}
\frac{\partial \bold{s}}{\partial t}=\frac{-J_{\text{ex}}}{\hbar m_{eq}}\bold{s}\times\bold{m}+R(t)\hat{n}-\frac{\bold{s}}{\tau_{\text{rec}}}
\end{eqnarray}
\end{center}
In Eq. (1), $\bold{m}$=$\vec{M}(T)$/$M_{S}^0$, with $M_{S}^0$ denoting the saturation magnetization at zero temperature and $\gamma$ is the gyromagnetic ratio. The first, second and third terms describe the precession, transverse relaxation and  longitudinal relaxtion of $\bold{m}$, respectively. $\bold{H}_{\text{eff}}$ is the effective field comprising internal anisotropy fields and the internal field related to longitudinal magnetization relaxation, the external geometrical pinning field and the applied field, and the exchange field including the field generating the oSTT.  The two  parameters $\alpha_{\perp}(T)$  and $\alpha_{||}(T)$ are the transverse and longitudinal damping parameters, respectively. (For more details see Supplementary information.) 

Eq. (2) describes the time-evolution of the spin polarized photo-electron density $ \bold{s}$. The first term is the precession of $ \bold{s}$ around the exchange-field of $ \bold{m}$ with the coupling strength $J_{\text{ex}}$; $m_{eq}$ is the normalized equilibrium magnetization. The second term describes the spin-polarised photo-electron injection rate $R(t)$, which is non-zero only during the $\sim100$~fs LP , and $\hat{n}$ is the helicity dependent spin-polarization. Depending on the light-helicity, $\hat{n}$ is $[0\text{ }0\pm1]$. The last term describes the decay of the spin density, limited primarily by the recombination time of the photo-electrons, $\tau_{\text{rec}}$. The oSTT from $\bold{s}$ on $\bold{m}$ is taken into account in Eq.(1) by adding the exchange field of $\bold{s}$ to $\bold{H}_{\text{eff}}$. 

In the simulations, we consider a Bloch DW subjected to LPs and the restoring field $B_{R}(x)$ as in Fig. 2B.  $B_{R}^{max}$ was set to a reduced value of 0.1 mT due to heat (deduced from Fig. 3B and described in the Supplementary information). 
Figs. 5 A,B show the simulated time evolution of $\bold{m}$ and $\bold{s}$ at the initial DW center during and after the application of a single 150 fs pulse with $\sigma^{+}$ ($~\hat{n}$=[0\text{ }0\text{ }1]) polarization. 

In Fig. 5A, the fast precession of $\bold{s}$  around the exchange field of $\bold{m}$ takes place until the photo-electrons recombine.  Only during this short time, angular momentum is transfered to $\bold{m}$. The precession of $\bold{s}$ is much faster than the  dynamics of $\bold{m}$ so that a significant change of $\bold{m}$ due to the precession around $\bold{H}_{\text{eff}}$  happens $\it{after}$ the photo electrons recombined. Fig 5B shows the time evolution of $\bold{m}$ at the center of the initial DW (${\bf m}$ is initially directed along $+\hat{\textbf{y}}$ for the Bloch DW).  During the short oSTT, $\bold{m}$ is only weakly disturbed from its equilibrium direction It takes $\sim1$~ns before it is rotated towards the $\hat{\bf z}$ axis.  At this time,  the center of the initial DW becomes part of the reversed domain and the DW has shifted by half width.
The deformation of the moving DW from the equilibrium Bloch DW profile is shown in Fig. 5C. The deformation $\Delta \bold{m}$ is obtained by subtracting the moving DW from the undisturbed Bloch DW profile after having shifted the center positions of the two DWs to $x=0$. Shortly after the LP exposure at $t=50$~ps, the DW magnetisation is strongly distorted. The simulation indicates that even after 5~ns, ${\Delta m_x(0)} \approx 0.15$ , so that the original Bloch DW is still deformed towards a N\'eel DW. The deformation of the DW from its equilibrium profile long time after the LP was applied causes magnetization precession around the arising effective fields and keeps the DW moving.
In Fig. 5D,  the  DW position versus time is plotted during the first three LPs. As can be seen, the entire DW moves predominantly between and not during the pulses. 

A calculation confirming the depinning of the DW from the cross is shown in Supplementary information. Here, oSTT pulses are applied until the DW reaches the cross center and overcomes the maximum value of the geometric pinning potential.  Our simulations fully confirm the experimental observations and the inferred picture in which the inertial motion is responsible for the DW displacement driven by the ultra-short LPs.

We finally remark that the helicity dependent DW motion can be also relized by a continuous light excitation. However, as confirmed in the Supplementary information both experimentally and by simulations the inertial DW motion driven by ultrashort LPs is more efficient.  
We also remark that the LP induced helicity dependent DW motion is not limited to diluted magnetic semiconductors. The oSTT induced DW motion may also be realized in heterostructures, where the spin-polarised photo-carrier excitation and spin transfer torque are spatially separated, e.g., when spin-polarized photo-electrons are injected from an optically active semiconductor into an adjacent thin ferromagnetic film. In this case, the oSTT can be equally efficient as found in our present study since the total magnetic moment of a $\sim 1$nm  thin magnetic transition metal film is comparable to the total magnetic moment of our  25~nm thick diluted magnetic semiconductor film with $\sim 5\%$ Mn doping. Indeed, the DW motion in a ferromagnetic film driven by spin-polarized currents applied electrically in the direction perpendicular to the film-plane has been recently proposed \cite{Khvalkovskiy_2009} and experimentally observed, showing a very fast DW motion \cite{Boone_2010} and low driving current densities \cite{Chanthbouala_2011}. 
Our concept represents an optical analogue of these electrical driven DW experiments with the potential of delivering orders of magnitude shorter while still highly efficient spin torque pulses. 


We acknowledge support from EU ERC Synergy Grant No. 610115, from the Ministry of Education of the Czech Republic Grant No. LM2011026, from the Grant Agency of the Czech Republic under Grant No. 14-37427G, and by the Grant Agency of Charles University in Prague Grants no. 1360313 and SVVÐ2015Ð260216.

\pagebreak

\begin{figure}[H]
\hspace*{-0cm}\includegraphics[scale=0.6]{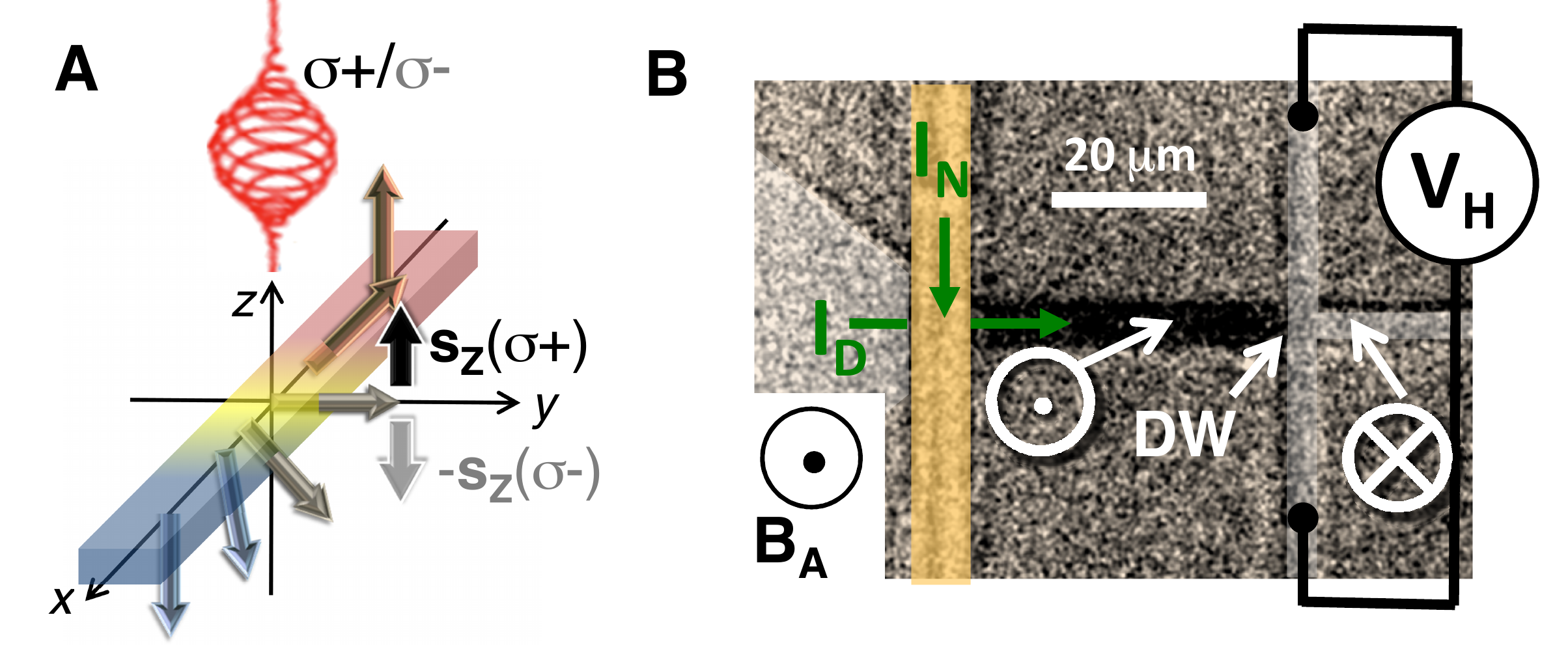}
\caption{(A) Sketch of light-helicity dependent optical spin transfer torque on a DW: Optically generated spin-polarised photoelectrons exert spin-transfer torque only on the rotating magnetization of the DW in the perpendicular magnetised film. Outside the DW, electron spin-polarization and magnetization are collinear.(B) Differential MOKE image of the initialized DW position at the cross entrance. After saturation, a reversed domain is nucleated by the Oersted field generated by the nucleation current $I_N$. 
Subsequently, the single DW propagates to its initial position when applying a small magnetic field of  $B_A \sim 0.2$mT. The initial straight DW position can also be detected by a AHE measurement  when applying the current $I_D$ along the Hall bar. The corresponding Hall signal $V_H$ corresponds to $\sim 11\%$ of the total signal upon compete magnetization reversal.
}

\label{f1}
\end{figure}

\begin{figure}[H]
\hspace*{-0cm}\includegraphics[scale=0.6]{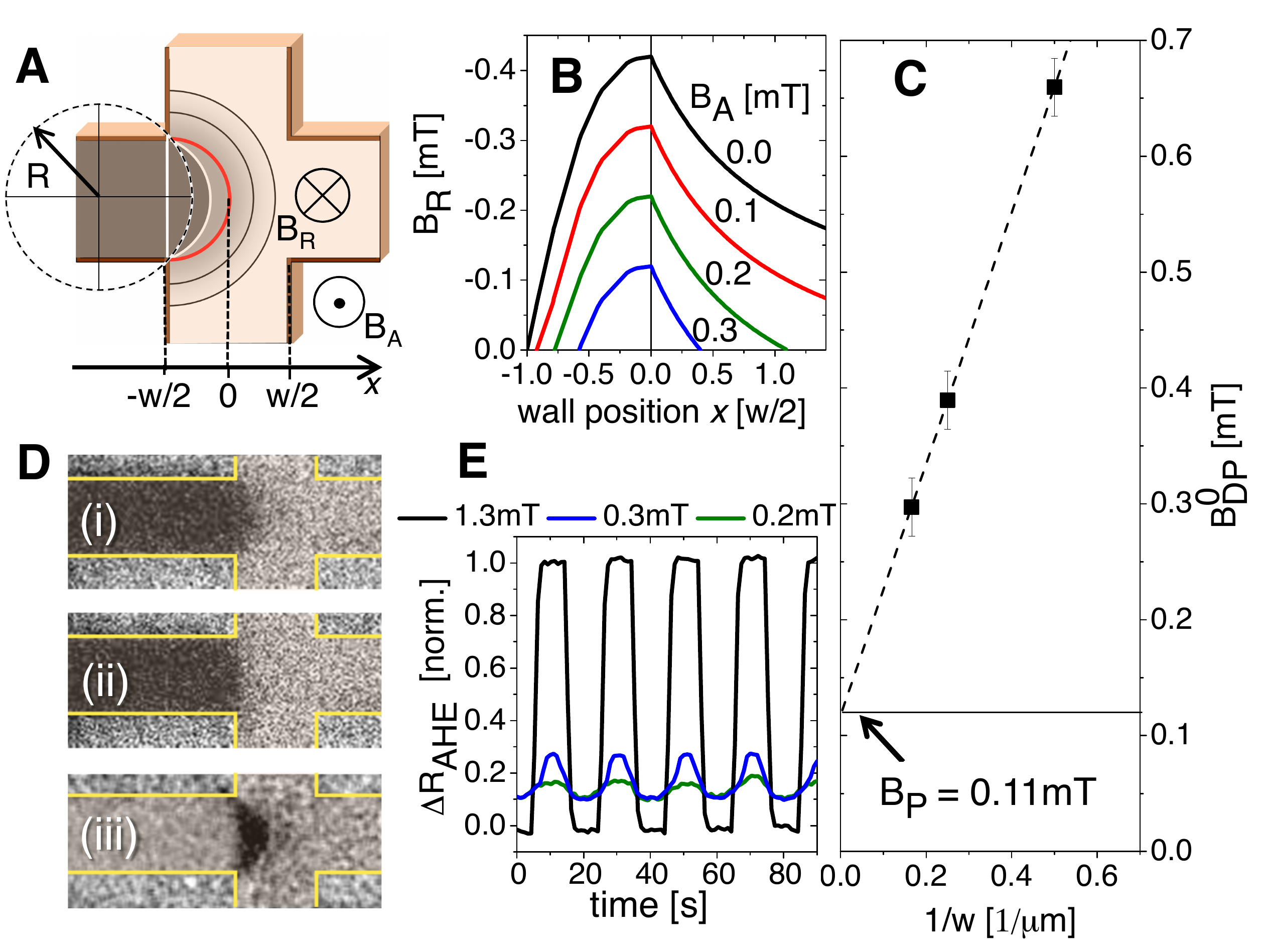}
\caption{(A) Schematic sketch of soap-bubble like extension of an elastic DW within a symmetrical cross under the application of a magnetic field: The domain wall stays pinned on the input corners until it reaches the cross center (red half-circle). (B) Effective restoring field $B_R$ opposing wall propagation at various applied magnetic fields $B_A$ as a function of DW position. $|B_R|$ becomes maximal when DW reaches the cross center. $B^0_{DP}$ corresponds to the lowest applied magnetic field necessary to depin the DW from the cross without LP irradiation and is equal to $|B_R^{max}| + B_P$, $B_P$ is the DW propagation field of the unpatterned magnetic film. (C) Experimentally determined depinning fields of 3 different devices with bar widths of $w = 2, 4$ and $6\mu$m. (D) Differential MOKE images of the $6\mu$m wide device at $B_A=0.25 $~mT (i) and at $B_A=0$~mT (ii). (iii): Bubble like domain shape when subtracting (ii) from (i). (E) Relative change of AHE signal (normalized to the total AHE signal upon compete magnetization reversal) due to elastic DW repulsive motion driven by an alternating field excitation $B_A =B_0|sin(w t)|$ with $B_0 = 0.2$mT, (green), $B_0 = 0.3$mT, (blue). AHE signal for complete magnetization reversal with $B_A=1.3$mT$\cdot sin(w t)$, (black). 
}

\label{f1}
\end{figure}

\begin{figure}[H]
\hspace*{-0cm}\includegraphics[scale=0.6]{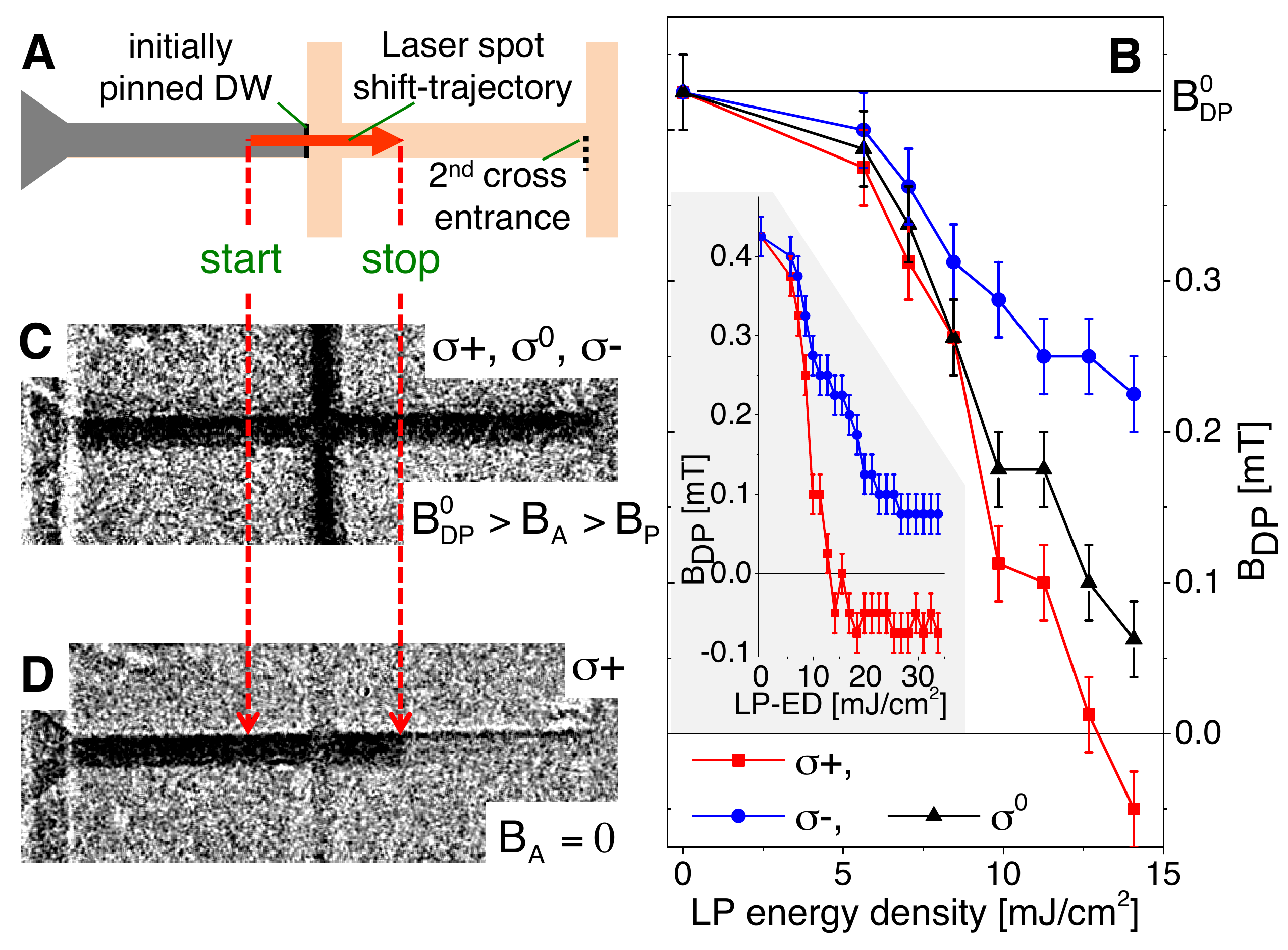}
\caption{(A) Experiment: To obtain $B_{DP}$, we first saturate the magnetization in a strong negative field. Then, a reversed domain is nucleated and a DW is positioned  at the cross entrance. The laser spot is now focused to its 'start' position 10$~\mu$m away from the initial DW location within the reversed domain. Subsequently, the spot is swept by 20$~\mu$m along the bar crossing the initial DW position with a velocity of $\sim2~\mu$m/ms. Starting from a small negative applied field  of $B_A$ =~-~0.1 mT, DW depinning is inferred from AHE measurements and differential MOKE micrographs taken after the laser spot sweep at constant $B_A$. If the DW is still located at the cross entrance, $B_A$ is increased by +0.025~mT followed by another laser spot sweep and subsequent AHE and MOKE detection. This procedure is repeated with stepwise increased $B_A$  until DW depinning is detected. Each individual data-point of $B_{DP}$ is obtained as the average from 5 independent depinning field measurements.
(B) Depinning field $B_{DP}$  as a function of LP energy density for circularly left (red), linearly (black) and circularly right (blue) polarized light. The inset shows $B_{DP}$  for circularly polarised light up to the highest LP energy density where the temperature increase due to LP heating does not  exceed the Curie temperature of the magnetic film. (Supplementary information,  )  
(C) Final domain configuration after laser sweeps with an applied field larger than propagation field $B_{P}$ and (D) at zero or small negative applied magnetic field.
}

\label{f1}
\end{figure}

\begin{figure}[H]
\hspace*{-0cm}\includegraphics[scale=0.6]{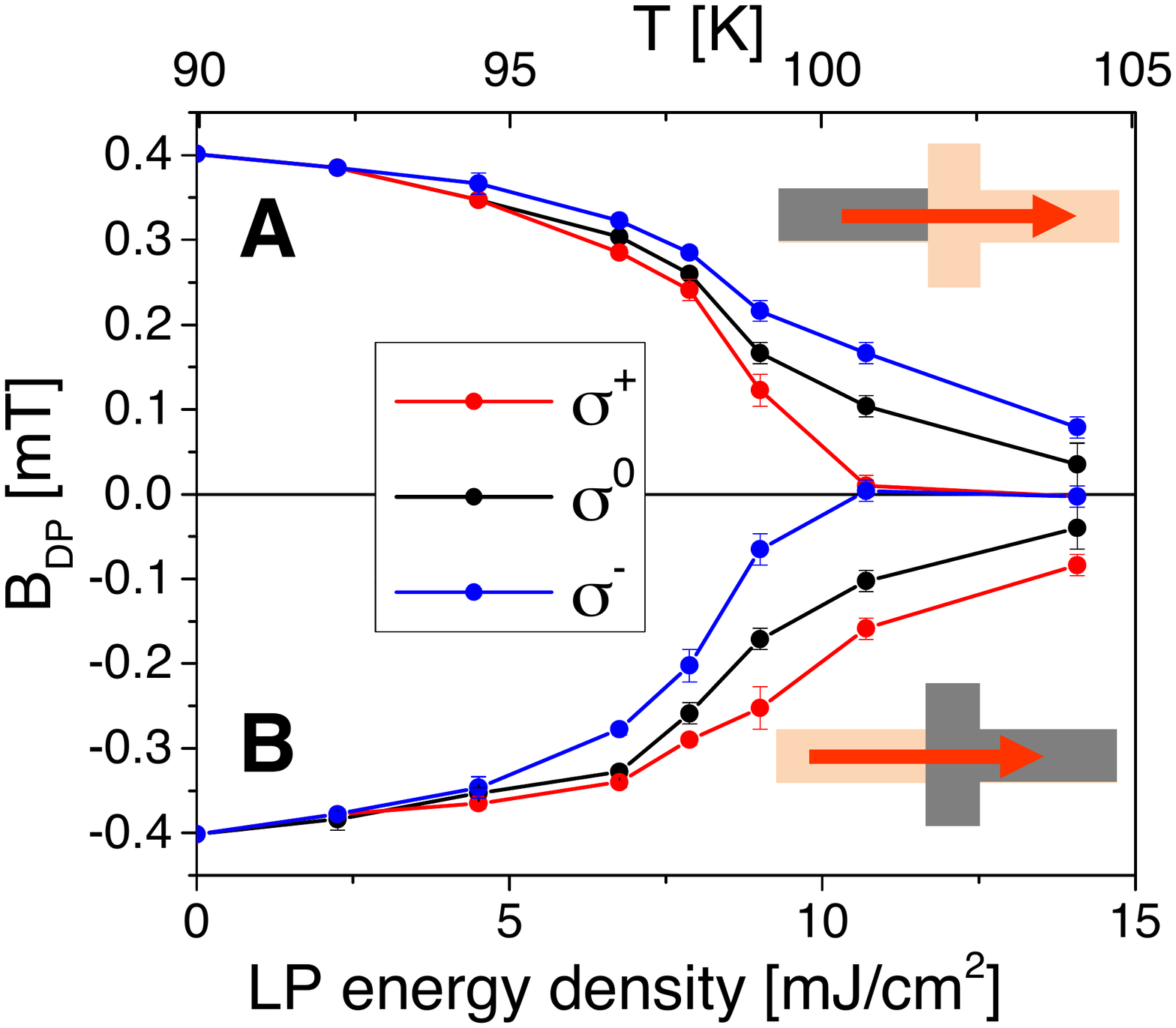}
\caption{Depinning field $B_{DP}$  as a function of LP energy density for circularly left (red), linearly (black) and circularly right (blue) polarized light with positive (A) and negative (B) nucleated domain magnetization. The LP related temperature increase estimated from the comparison between $B_{DP}$($\sigma^{0}, T = 90$~K) and  $B_{DP}$($0,~T$) is plotted at the top of the graph. 
}

\label{f1}
\end{figure} 
 
\begin{figure}[H]
\hspace*{-0cm}\includegraphics[scale=0.6]{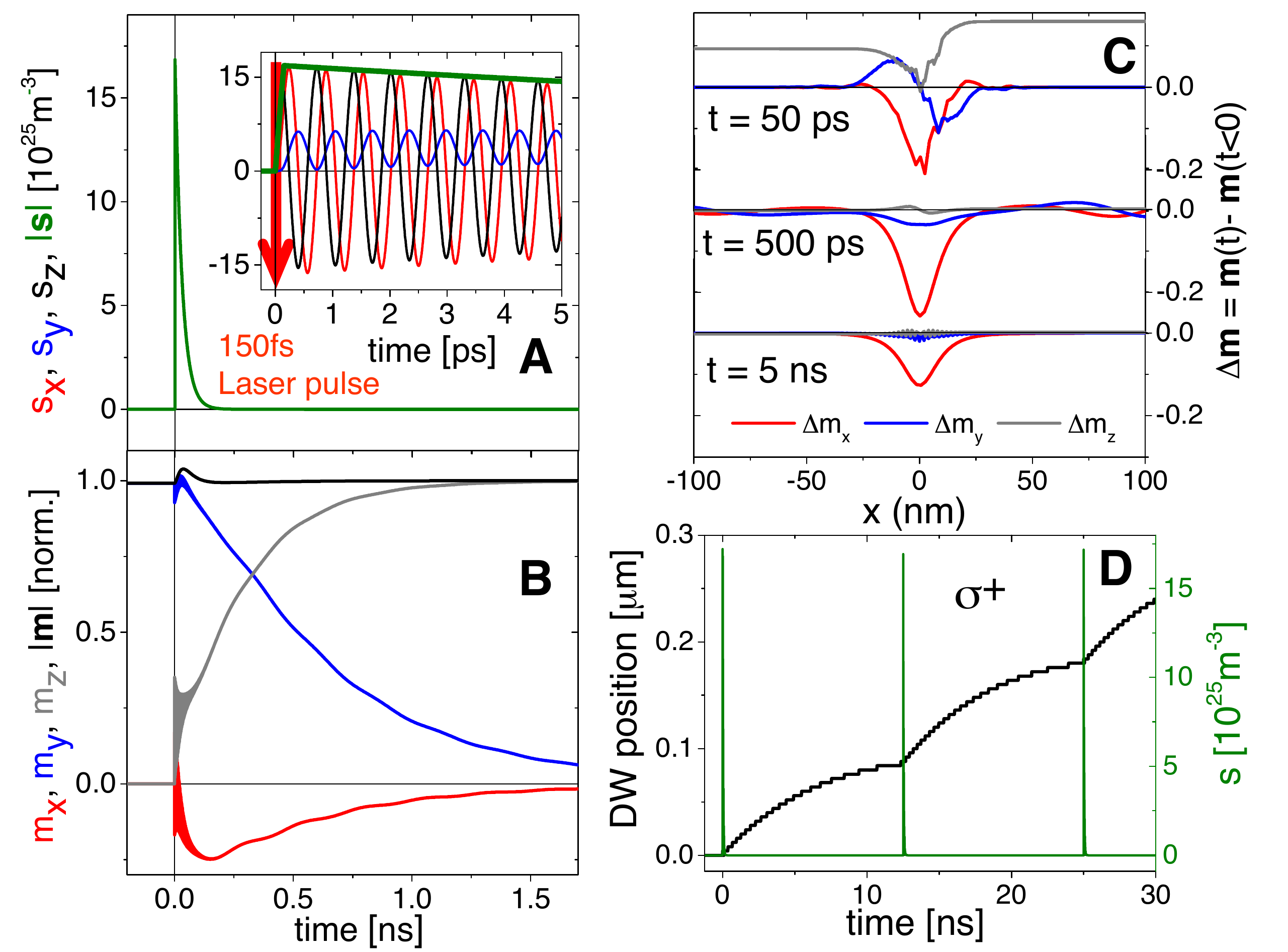}
\caption{(A) Simulated time evolution of photo-electron spin density $|\bold{s}|$ at the center of the DW generated by a150~fs long LP. Inset: The $x,y,z$-components of $\bold{s}$  vs. time $t$ showing the fast precession around the exchange field of magnetization $\bold{m}$. The red arrow in the inset indicates the  LP. (B) The components of $\bold{m}$ and $|\bold{m}|$ vs. $t$ at a fixed position corresponding to the initial DW center. At $~t=0$, $\bold{m}$ is oriented along the $y$-direction at the center of the Bloch-like DW. Note that $\bold{m}$ has been normalized by its modulus before the LP is applied. The graph shows a fast initial excitation due to the LP and a damped fast jiggling during the recombination time of the photo-electrons. During this short time, angular momentum is transferred from $\bold{s}$ to $\bold{m}$ causing a deformation of the DW. Note,  that during the oSTT, the magnitude of  $|\bold{m}|$ increases due to the interaction between the non-zero $y$-component of the precessing spin density and the magnetisation at the DW center oriented initially also along $y$. (C) Time evolution of the DW deformation $\Delta\bold{m}$. The 3 plots show the time-evolution of the deviation from the undisturbed DW profile in the rest frame of the domain wall with zero at the DW center after the pulse was applied. The slowly relaxing DW deformation causes the DW motion. (D) The DW position as a function of time for the first three  $\sigma^{+}$  polarized LPs (a pulse occurs every 12.5 ns).
In (A)-(D), $\tau_{rec}=30$~ps and $R=\text{1.2}\times\text{10}^{39}\text{m}^{-3}\text{s}^{-1}$.
}

\label{f1}
\end{figure}

\newpage

 \pagebreak
 
 \title{Supplementary Material: \\ Inertial displacement of a domain wall excited by ultra-short circularly polarized laser pulses}

\maketitle 

\section{Elastic bubble expansion and geometrical pinning}
To describe the geometrical DW pinning exploited in our experiment we employ a simple DW propagation model where the motion of a DW of negligible width is determined by the competition between DW energy   $E_{\sigma}=\sigma \cdot t \cdot l$ and Zeeman energy $E_Z=-2 M_s \cdot H_a  \cdot t \cdot S$.  $\sigma = 4 \sqrt{A  K_{E}}$ is the DW energy per unit area, $M_s$, $A$, and $K_{E}$ are saturation magnetisation, exchange stiffness and effective uniaxial anisotropy constants, respectively, $l$ is the DW length, $S$ is the area of the reversed domain and $t$ is the thickness of the magnetic layer. The 'friction' of DW motion generated by DW pinning on defects in the magnetic film is described in our model by a coercive \textit{intrinsic} propagation field $H_p$ which is considered to be everywhere the same. We also neglect the effect of magnetic pseudo-charges generated by DW deformation, since the radius of curvature is much larger than the width of the DW in our micrometer wide Hall-crossbar structures. We therefore consider magnetic field driven DW propagation to be governed only by the competition between DW energy and Zeeman energy. 

We first consider a circular-shaped domain of radius $r$ expanding around a nucleation center in a magnetic plate without any geometrical restrictions. Minimizing the total energy  $E_{tot}=E_{\sigma}+E_Z$ yields  a DW with minimal length bordering a maximal area of the reversed domain. If the wall propagates by $dq$, the total energy changes by 
\begin{align}
\frac{d E_{tot}}{d q}=\sigma t \frac{d l}{d q} - 2 M_s H_a t \frac{d S}{d q} \equiv 2M_S(H_R-H_a)2\pi rt \text{,}
\end{align}

where  ${H_R=(2M_s)^{-1}\sigma/r}$ is hereafter defined as the \textit{virtual} restoring field. The effective net field $H_{net}$ oriented along the layer normal direction, which induced the DW propagation, contains in addition to the applied field $H_a$, the propagation field $H_p$ and the damping torque field  $H_{\alpha} $  \cite{Malozemoff79}  as well as the contribution $H_R$ arising from the DW curvature \cite{Wunderlich2001}. 

In our geometrically constricted Hall-crossbar of width $w$, we can also reduce the DW motion to a one dimensional problem, that of a virtually straight DW propagating at the position $q$ with the velocity $v$ of the real DW center, and on which a restoring force acts due to the curvature of the real wall.
Within the bar outside of the Hall cross, the DW of $l=w$ stays straight perpendicular to the stripe boundaries, and propagates in the applied magnetic field. When the DW reaches the two corners of the cross entrance, the DW must increase its length to continue propagation. The energetically optimal way is a bubble like expansion as shown in Fig. S1 where the DW starts as a flat line (stage 'B') and remains connected to the corners of the cross entrance until it coincides with a semi-circle of radius $w/2\equiv d$,  (stage 'C'). During this process where the DW propagates from the position $q=-d \rightarrow 0$,  the DW length enhances as $l=2r \cdot \text{arcsin}(d/r)$ and the reversed domain surface increases as $S=r^2 \Big[\text{arcsin}(d/r)-(d/r) \sqrt{1-(d/r)^2}\Big] $ since the DW curvature radius shrinks from $r=\infty \rightarrow d$. The relation between $r$ and $q$ is given by $q=r-d-\sqrt{r^2 - d^2}$. The total energy variation is given by 
\begin{align*}
\frac{d E_{tot}}{d q}= ( \sigma - 2 M_s H_a r ) t\frac{\sqrt{r^2 - d^2}\text{arcsin}(d/r)-d}{\sqrt{r^2 - d^2} - r} \text{.}
\end{align*}
Beyond the semi-circle (stage 'C'-'D'), the two entrance corners do no longer influence the DW propagation and the domain continues to expand circularly with increasing radius. The relation between curvature radius $r$  and propagation coordinate $q$ is now $r=q+d$ and the energy variation leads to
\begin{align*}
\frac{d E_{tot}}{d q}= ( \sigma -  2 M_s H_a r ) t \pi  \text{.}
\end{align*}

The equilibrium conditions $d E_{tot}/d q=0$ leads to the applied magnetic field at which restoring and driving forces are balanced: $H_R(q)=\frac{\sigma}{2M_s\cdot r'} $  within the range of  $q = -w/2  \rightarrow 0$ , $q=r'-d-\sqrt{r'^2 - d^2}$ and $H_R(q)=\frac{\sigma}{2M_s [q+d]} $  within the range of  $q = 0 \rightarrow (\sqrt{5}-1)d$,  hence, the maximum of the restoring field 
\begin{align}
H_R^{\text{max}}=\frac{\sigma}{M_s w} 
\end{align}
is reached, when the DW reaches the cross center at $q=0$. 

According to our model, both virtual restoring field $H_R$ and the intrinsic propagation field $H_p$ oppose the DW propagation. Thus, DW propagation without thermal activation will start as soon as the applied field exceeds the value of $H_R + H_p$. In contrast to $H_p$, the restoring field $H_R(q)$ is a function of DW position and the model predicts a maximum depinning field at the center of the cross. Since $H_R^{\text{max}}$ is inverse proportional to the bar width, larger pinning in narrower bars is expected. 
Indeed, our experimental findings shown in the main text (Fig. 2) are in very good agreement with the theoretical pretictions of our model and confirm that the geometrical pinning dominates by far the intrinsic pinning and also possible pinning at the pair of exit corners (stage 'D') due to demagnetisation field inhomogeneities  \cite{Wunderlich2001}. 

In the same spirit, virtual restoring field and intrinsic pinning on defects oppose DW motion when the driving mechanism is optical spin-transfer torque generated by circularly polarised laser pulses (LPÕs). This allows us to relate a measure of the oSTT driven domain wall motion to the geometrical pinning strength which is always measured as a reference quantity in a field assisted depinning experiment without irradiation. 
In Fig. S2, we present data of polarisation  dependent depinning fields $B_{dp}$ of 2, 4 and 6~$\mu$m wide crossbar devices as a function of LP energy density. Without laser irradiation, $B_{dp}$ is largest for the narrowest bar. Thermal heating by the LP irradiation, however, is more effective for the narrower device so that the reduction of $B_{dp}$ is faster with increasing LP energy density for the narrower bars.
	
\section{Micromagnetic method}
\subsection{Landau-Lifshitz-Bloch approach (LLB)} 
We state again here firstly the governing equations as the forthcoming description is centered around them: The time evolution of the magnetization $\vec{m}$ in the LLB approach ~\cite{Schieback} and the spin-density $\vec{s}$ ~\cite{Nemec} read :
\begin{eqnarray}
\frac{\partial\vec{m}}{\partial t}=-\gamma\vec{m}\times\vec{H}_{\text{eff}}+\vec{\Gamma}_{\text{tr}}+\vec{\Gamma}_{\text{lt}} \\
\frac{\partial \vec{s}}{\partial t}=\frac{-J_{\text{ex}}}{\hbar m_{eq}}\vec{s}\times\vec{m}+R(t)\hat{n}-\frac{\vec{s}}{\tau_{\text{rec}}}
\end{eqnarray}

We start by describing Eq.(3): There, $\vec{m}$ is the magnetization at temperature $T$, normalized by the zero temperature saturation magnetization $M_{0}$ , $\gamma$ is the gyromagnetic ratio and $\vec{H}_{\text{eff}}$ is the effective field (as described later on). $M_{0}$ was determined by extrapolation of SQUID-data and is here 35.5 kA/m.  The first term on the right hand-side describes the precession of $\vec{m}$ around $\vec{H}_{\text{eff}}$ while the second term $\vec{\Gamma_{\text{tr}}}$=$-\frac{\gamma\alpha_{\perp}}{m^{2}}\vec{m}\times\left(\vec{m}\times\vec{H}_{\text{eff}}\right)$ is the transverse torque with associated damping $\alpha_{\perp}(T)$=$\lambda\left(1-\frac{T}{3T_{C}}\right)$, resulting in relaxation of $\vec{m}$ into the direction of $\vec{H}_{\text{eff}}$. Here, $\lambda$ is the microscopic damping parameter at $T$=0.
The first and second terms on the right hand-side of Eq.(3) consitute the torques included in the LLG description. In the LLB equation, a third term, $\vec{\Gamma}_{\text{lt}}$=$\frac{\gamma\alpha_{||}}{m^{2}}\left(\vec{m}\cdot\vec{H}_{\text{eff}}\right)\vec{m}$ is present, allowing for a longitudinal variation of $\vec{m}$; in other words $|\vec{m}|$ is not conserved and is allowed to fluctuate with an associated damping parameter $\alpha_{||}(T)$=$\frac{2T\lambda}{3T_{C}}$ beacuse, at elevated temperatures, all atomic spins whose ensemble form the corresponding $\vec{m}$ in a computational cell, are not necessarily all parallel to each other at all times (which is the assumption and a constraint in LLG-micromagnetics).  Further, the interaction terms taken into account here result in $\vec{H}_{\text{eff}}$=$\vec{H}_{\text{d}}+\vec{H}_{\text{ex}}+\vec{H}_{\text{mf}}+\vec{H}_{\text{k}}+\vec{H}_{\text{OSTT}}+\vec{H}_{r}$, which are, demagnetizing, exchange, internal material field, uniaxial magnetocrystalline anisotropy, optical spin transfer torque and geometrical pinning -fields, respectively. The effective field terms are evaluated from the free energy density $f$ as $\frac{-1}{\mu_{0}M_{0}}\frac{\delta f}{\delta \vec{m}}$. $\vec{H}_{r}(x)$ is taken directly from Fig. 2 B  in the main text. Its temperature dependence is described in terms of the thermodynamic equilibrium functions of the pertinent material parameters; normalized  equilibrium magnetization $m_{eq}$ at a given $T$ (normalized by $M_{0}$), exchange stiffness $A(T)$, uniaxial magnetocrystalline anisotropy, $K_{\perp}(T)$, $K_{||}(T)$  and longitudinal susceptibility $\chi_{||}(T)$. Here, the temperature dependence of $m_{eq}$ was evaluated within the mean field approximation, by a Langevin function fit to measured SQUID-data and $\chi_{||}(T)$ was calculated as shown below.  
The equilibrium magnetization at $T=90 $~K is here 18 kA$\text{m}^{-1}$. $K_{||}$ at the temperature used in the simulations was estimated from data presented in De Ranieri et al.  \cite{DeRanieri2013} taken on nominally identical GaMnAsP material. The mean values of $K_{||}$=350 J$\text{m}^{-3}$ and $K_{z}$=1.51 kJ$\text{m}^{-3}$ were determined from characterization measurements on single bar-devices at a temperature of $T=90$~K, and the exchange stiffness constant of $A(T=90$~K$)=50 ~\text{fJm}^{-1}$ is reasonable for Ga$_{0.94}$Mn$_{0.06}$As$_{0.9}$. Implementing the values above  in our bubble like DW propagation model reproduces the measured depinning fields at various bar widths. By using previously measured values valid for this temperature we avoid mean-field fitting for most material parameters and thus we are more certain of their realistic values.  For  dynamical simulations we choose the damping parameter $\lambda$=0.01. All simulations are performed considering a base temperature of $T=90$~K (in accordance to the experiments). 
The demagnetizing field is divided into the near-field and the far-field and is described in terms of the demagnetizing tensor, $\hat{N}$ in the standard manner; the dipole field at point $\vec{r}_{i}$ from all dipoles at points $\vec{r}_{j}$ is $\vec{H}_{d}^{i}(T)$=$-m_{eq}M_{0}\sum_{j}\hat{N}(\vec{r}_{i}-\vec{r}_{j},\Delta_{x},\Delta_{y},\Delta_{z})\vec{m}_{j}$, where $\Delta_{x,y,z}$ are the dimensions of the discretization cells used along $x,y$ and $z$, respectively. For the near-field, $\hat{N}$ is evaluated by the analytcial formulae for interactions between tetragonal cells as derived by Newell, Williams and Dunlop ~\cite{Dunlop}. For the far-field (here, for inter-cell distances $\geq$ 40 cells ), the kernel elements of $\hat{N}$ correspond to those for point dipoles. $\hat{N}$ need only be computed once and stored in memory. The form of $\vec{H}_{d}^{i}(T)$ is that of a spatial convolution. This convolution is then evaluated by standard FFT-techniques.  The exchange field $\vec{H}_{\text{ex}}(T)$=$\frac{2A(T)}{\mu_{0}m_{eq}^{2}M_{0}}\partial_{\vec{r}}^2\vec{m}$, where the second derivative is computed by a finite difference three-point stencil in each spatial direction. $\vec{H}_{\text{mf}}$, responsible for stabilizing $|\vec{m}|$ is determined by the parallel susceptibility $\chi_{||}(T)$ as $\vec{H}_{\text{mf}}(T)$=$\frac{1}{2\chi_{||}(T)}\left(1-\frac{m^{2}}{m_{\text{eq}}^{2}}\right)\vec{m}$ ($\chi_{||}=(\partial m_{\text{eq}}/\partial H)_{H\rightarrow0}$, with $H$ being an applied field). In this work, the global easy axis $\hat{u}||\hat{z}$ and the in-plane uniaxial anisotropy axis $\hat{u}||\hat{y}$. Each anisotropy term contributes to $\vec{H}_{k}(T)$ as $\vec{H}_{k}(T)$=$\frac{2K(T)}{\mu_{0}m_{eq}^{2}(T)M_{0}}(\vec{m}\cdot\hat{u})\hat{u}$. The last effective field term, $\vec{H}_{r}(T)$ is based on  considered pinning field profile for a bubble domain pinned at a cross, while the effect of temperature is taken into account by considering the reduction of $B_{dp}$ for $\sigma^{0}$-light at the laser fluency corresponding to DW depinning for $\sigma^{-}$-light. Therefore, the maximum  $|\vec{H}_{r}(T)|$ used in the simulation corresponds to 0.1 mT. As it acts as to pull the DW in the opposite direction of its excited motion, then in the simulations,  the direction of the virtual restoring field $\vec{H}_{r}(x)$ is along the $\mp z$-direction if the DW moves along the $\pm x$-direction. Finally, the boundary condition used for $\vec{m}$ on all free surfaces is $\frac{\partial\vec{m}}{\partial\hat{r}_{n}}=0$, where $\hat{r}_{n}$ is the outward unit normal.

We now turn to Eq.(4)  and its coupling to Eq.(3). Here, $J_{\text{ex}}$ is the exchange coupling between photo-induced electrons and the local magnetization $\vec{m}$. We use $J_{\text{ex}}$=$JS_{\text{Mn}}c_{\text{Mn}}$, where $J$=10 meV$\text{nm}^{3}$, $S_{\text{Mn}}$=5/2 is the local Mn-moment and $c_{\text{Mn}}\sim$ 1 $\text{nm}^{-3}$ is the typical moment density \cite{Nemec}. When coupling to the LLB-equation we assume a temperature variation of the effective exchange coupling to the macro-vector $\vec{m}$ at increased $T$ and  for simplicity assume $J_{\text{ex}}\rightarrow J_{\text{ex}}m_{\text{eq}}^{2}$. The first term on the right hand-side of Eq.(4) describes the precession of $\vec{s}$ around the exchange field produced by $\vec{m}$ (in this step the effect of $\vec{m}$ on $\vec{s}$ is established) while the second term gives the injection of spin-polararized electrons with $R$ being the rate per unit volume and $\hat{n}$ the initial spin polarization direction defined by the helicity of the light with $\hat{n}$=[00$\pm$1]. Finally, the third term represents the decay of the photo-carrier spin with a life-time of $\tau_{\text{rec}}$, limited in our case by the carrier-recombination time. Based on previous measurements in literature, we set $\tau_{\text{rec}}$=30 ps.   Gradient terms in $\vec{s}$ are neglected. During precession, $\vec{s}$ transfer its angular momentum to $\vec{m}$. The precession time of $\vec{s}$ is very fast as compared to the natural precession of $\vec{m}$ (100s of fs versus a few ns). The absorption of angular momentum from $\vec{s}$ results in a torque on $\vec{m}$. This torque is then entered into Eq.(3) by an augmentation to the rest of the effective field by $\vec{H}_{\text{OSTT}}(T)$ (thus the effect of $\vec{s}$ on $\vec{m}$ is established); The interaction energy density between $\vec{s}$ and $\vec{m}$ is $f_{ex}$=$\frac{-J_{\text{eff}}(T)}{m_{\text{eq}}}\vec{s}\cdot\vec{m}$. The corresponding effective field term is then according to the definition in the preceeding paragraph, $\vec{H}_{\text{OSTT}}(T)$=$\frac{J_{\text{eff}}(T)}{\mu_{0} m_{\text{eq}}M_{0}}\vec{s}$. Equations (3) and (4) are solved together using a $5^{th}$ order Runge-Kutta integration scheme.

\subsection{Computational geometry and simulation procedure}

We consider a one-dimensional bar with 4095 x 1 x 1 computational cells composing a structure as shown in Fig. S3. The cell dimension is 4 nm x 4 $\mu\text{m}$ x 25 nm.
A Bloch DW is initialized in the centre of the bar and let to relax quickly with strong damping by setting $\lambda$=0.9.
This configuration is then used as a starting configuration for the simulations of domain wall motion under the light pulses.

Once the domain wall is prepared, circularly polarized light is pulsed at a rate of 80 MHz. The length of each pulse is set to 150 fs. For the simulation of the depinning process, the spin-polarized carrier injection rate is $R= 1.225\times \text{10}^{39} \text{m}^{-3}\text{s}^{-1}$. This order of magnitude for $R$ is required that the DW can escape the elastic pinning potential. The equivalent pulse power corresponds to the time-averaged laser power used in the experiments of the order of $20$~mW assuming a skin depth of $1 ~\mu$m. Further, all simulations were done in zero externally applied magnetic field and a (zero temperature) damping of $\lambda= 0.01$ was used in all dynamical simulations. 

Throughout all simulations a centering procedure is employed, that keeps the DW in the middle of the length of the bar. In this way, propagation distances as long as needed can be simulated without having to worry about stray field effects should the domain wall have come close to the edges of the bar or that the DW moves out of the computational region.

\section{Additional experimental evidences for optical spin transfer torque driven DW motion} 
\subsection{Wavelength dependency of helicity dependent DW motion}
We have performed wavelength dependent experiments to support the optical spin transfer torque origin of the helicity dependent LP induced DW motion. 
In experiments described in the main text we use LP excitation with  a wavelength $\lambda$ = 750~nm that excites photo-electrons slightly above the bottom of the GaAs conduction band so that for a circularly polarized incident light, photo-electrons become spin-polarized with the degree of polarization approaching the maximum theoretical value of $50\%$ \cite{Pikus1984}. At energies above and below band gap energy (Fig. S4), the spin polarisation of the photo-electrons is reduced. Photons excitation at higher energies results in the reduction of net-photoelectron spin-polarisation  mainly because carrier excitation from the  split-off valence band can take place. The photocarrier generation from low-energy photons with sub-band gap energies comes mainly from excitation of impurity states within the band gap.

\subsection{Dependency of helicity dependent DW motion on the sweep direction of the focused laser spot}
We now investigate the effect of the thermal gradients generated by the laser spot on the helicity dependent DW depinning. By inverting the sweep direction of the focused LP spot with respect to the geometrically pinned DW, we invert also the thermal gradient affecting the DW. In case that the LP spot approaches from the reversed domain along the patterned bar, Fig. S5A, both $B_{dp}(\sigma ^+)$ and $B_{dp}(\sigma ^-)$ decrease faster compared to the situation where the laser spot approaches from the unreversed domain (Fig. S5B). This observation is explained by the stronger temperature increase from LP heating in the narrow bar compared to the lower temperature rise in the wider cross area. On the other hand, the helicity dependence of $B_{dp}$, which is of oSTT origin, remains unaffected by the LP sweep direction.

\subsection{Continuous wave excitation vs. laser pulse excitation}
We now show that the DW can be also moved by a focused laser spot of continuous wave (cw) excitation. To compare the efficiency of the cw-excitation with our ultrashort LP-excitation approach we have deduced  $B_{dp}(\sigma ^+)$ and $B_{dp}(\sigma ^-)$  for LP- and cw-excitation at the same averaged laser power $P_{av}$. Based on our LLB approach, we also have calculated DW propagation driven by oSTT from LP- and cw- excitations. From the simulations, we have derived the averaged DW velocity vs. $P_{av}$ at the position of maximal restoring field (Fig. S6A). Positive averaged velocities correspond therefore to the depinning of the DW. For zero or negative velocities, the DW remains pinned. Both calculation and experimental results presented in Fig. S6 confirm that a DW can be depinned via oSTT generated by ultrashort LP and by cw-excitation. However, cw-driven DW propagation requires always higher averaged laser power (Fig. S6A). Comparing $B_{dp}(\sigma ^-, \sigma ^+)$ (Fig. S6B, C) for the two excitation schemes at equal averaged laser power shows a stronger efficiency of the oSTT. Moreover, the effect of laser heating on DW depinning is stronger in case of cw-excitation compared to ultrashort LP-excitation. Therefore, helicity dependent inertial DW motion induced by ultrashort LP is more efficient than DW motion induced by constant excitation.

\subsection{Temperature dependent depinning field}
The temperature dependence of the resistance of the magnetic bar is used to monitor and control the actual sample temperature (Fig. S7A).
In order to obtain the accurate resistivity dependence of our devices we performed a reference measurement in a bath cryostat, where the (Ga,Mn)(As,P) sample is thermally anchored to a calibrated temperature sensor and where the temperature dependent resistivity $R(T )$ of (Ga,Mn)(As,P) is monitored during heating-up from $4~$K to room temperature. The Curie temperature $Tc = 115$~K is obtained by identifying the cusp in $dR/dT$ ~\cite{Novak08}, (Fig. S7A).
We have determined $B_{dp}$ without laser irradiation at a $4~\mu$m wide device as a function of temperature in a temperature range below $T=90$~K until close to Curie-temperture (Fig. S7B). This allows us to estimate an effective sample temperature deduced from the comparison between temperature dependent measurements of $B_{dp}$ without irradiation and measurements of $B_{dp}(\sigma ^0)$ vs. $P_{av}$  laser power and at fixed base temperature \cite{Shihab2016}.

\pagebreak
\setcounter{figure}{0}
\makeatletter 
\renewcommand{\thefigure}{S\arabic{figure}}

\begin{figure}[H]
\hspace*{1cm}\includegraphics[scale=0.55]{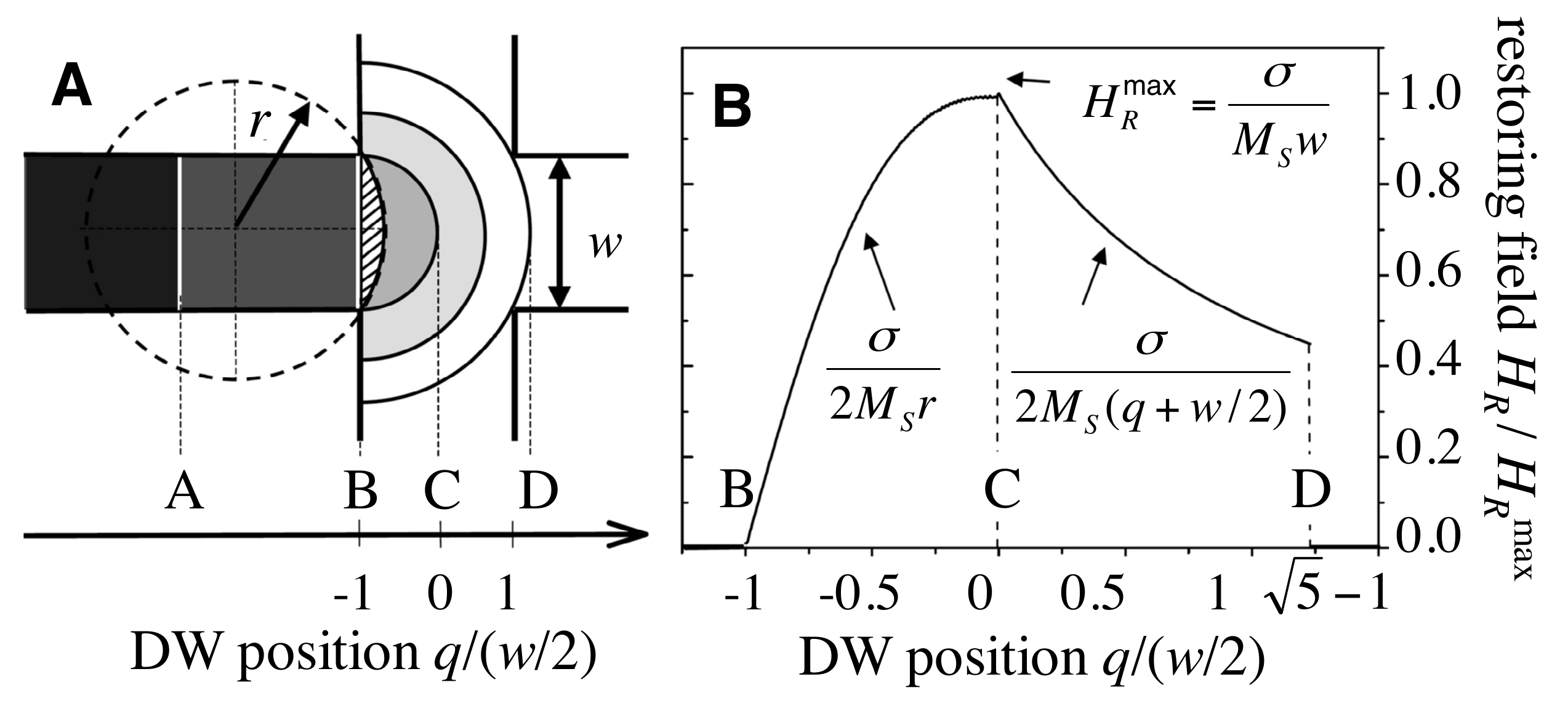}
\caption{(A) "Soap-bubble" like expansion of a DW within a symmetric cross. The DW stays pinned on the cross input corners until stage 'C' is reached. At this position, the geometrical restoring field $H_R$ reaches its maximum.
(B) Position dependent \textit{virtual} restoring field $H_R(q)$ arising from the wall curvature is reflecting the elasticity of the wall. Introducing $H_R(q)$ reduces our system to a one-dimensional problem, that of a virtual straight DW propagating with position  $q$  and velocity $v$ of the real DW center.}
\end{figure}

\begin{figure}[H]
\hspace*{1cm}\includegraphics[scale=0.55]{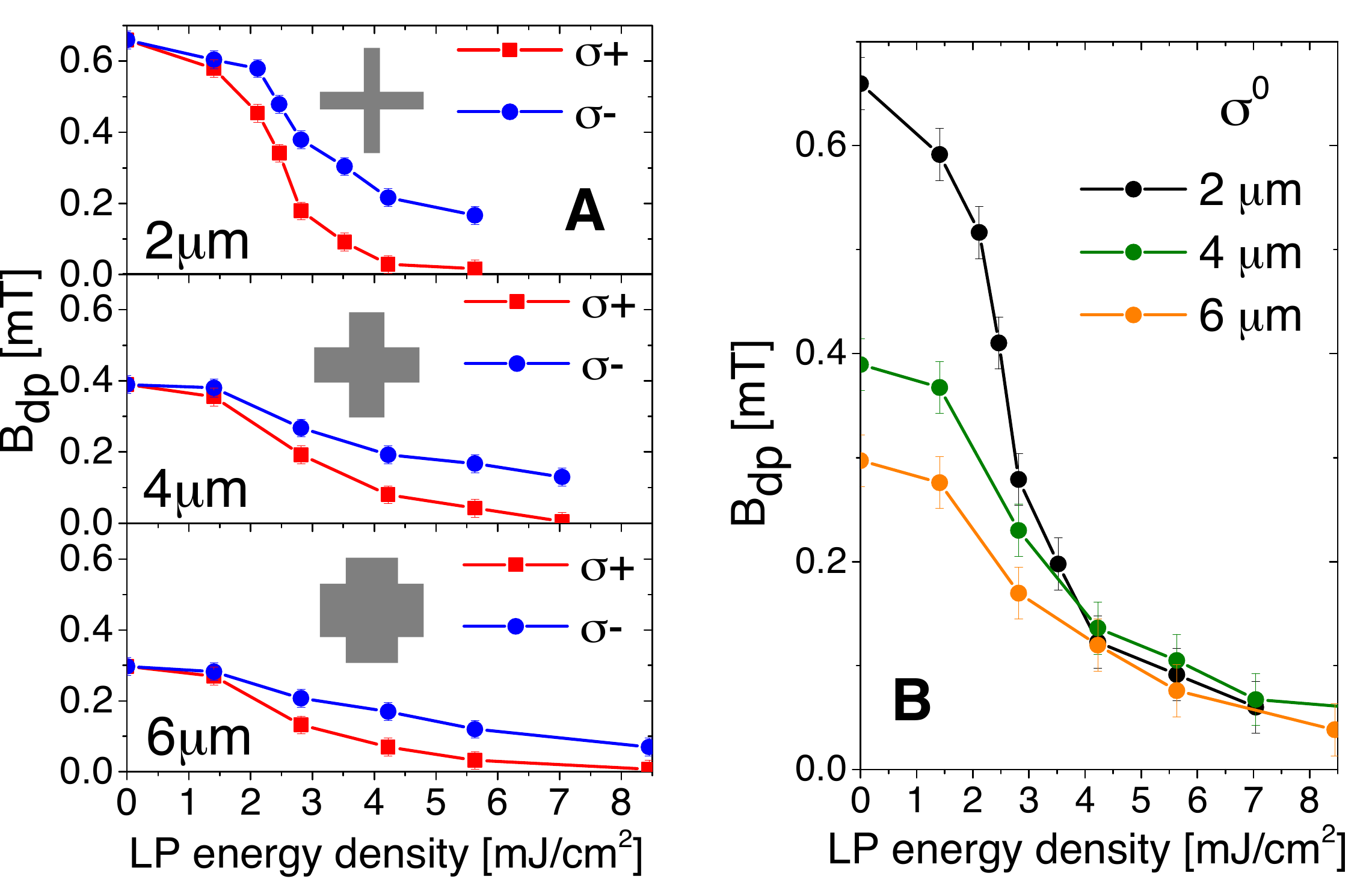}
\caption{Polarization dependent depinning field  $B_{dp}(\sigma ^+, \sigma ^-, \sigma ^0)$ as a function of LP energy density of 2, 4 and 6~$\mu$m wide crossbars for (A) circularly ($\sigma ^{+,-}$) and (B) linearly ($\sigma ^0$) polarized LPs. The difference between $B_{dp}(\sigma ^+)$  and $B_{dp}(\sigma ^-)$ is due to the optical spin transfer torque and the decrease of $B_{dp}$ with increaseing LP energy density is  due to the heating from photon absorption.}
\end{figure}
	
\begin{figure}[H]
\hspace*{-1cm}\includegraphics[scale=0.55]{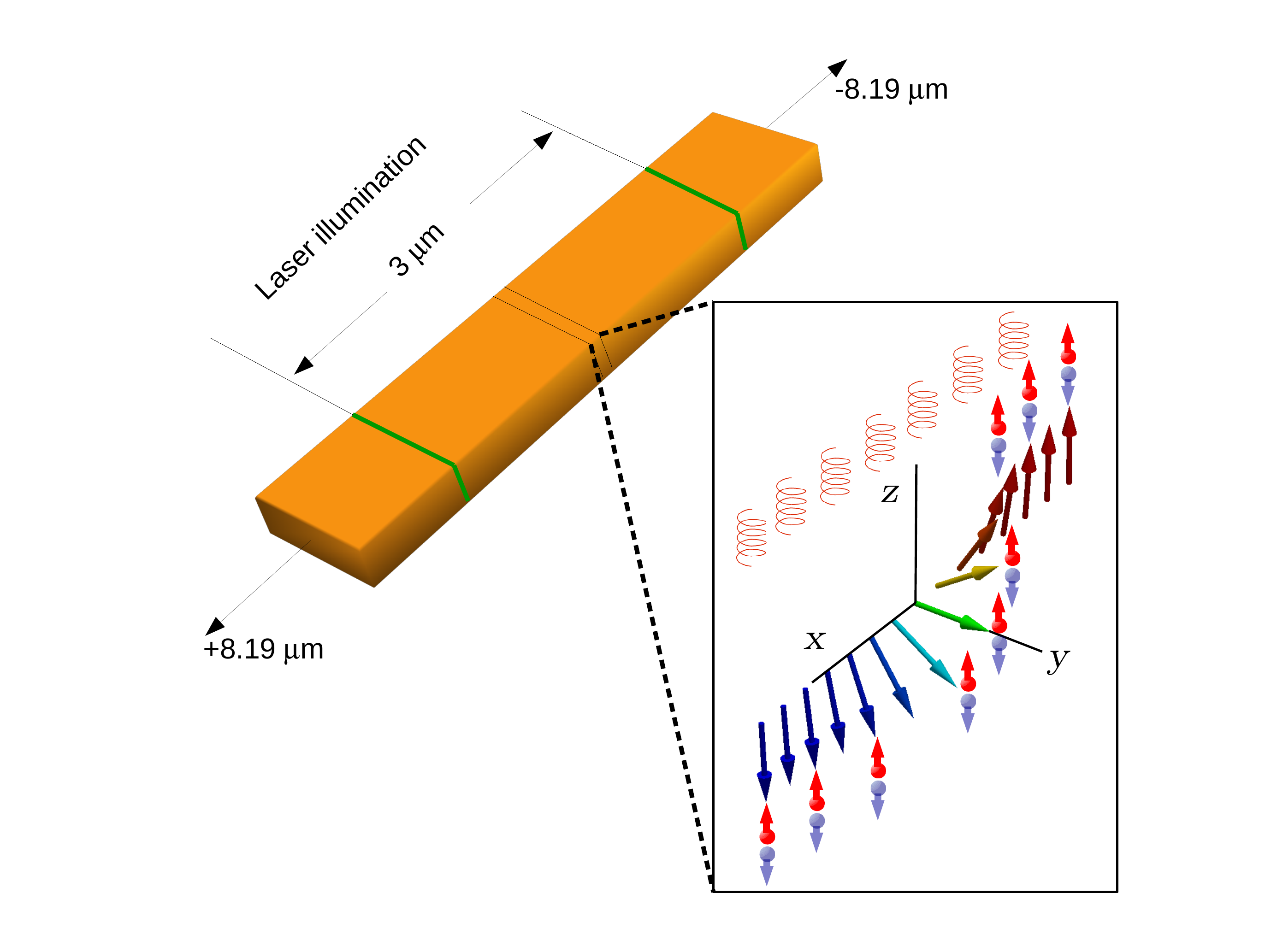}
\caption{Computational Setup: The total bar extension is 4x4095=16384 nm along $x$, 4 $\mu$m wide and 25 nm thick. The DW is of Bloch type, initially located at the centre of the bar. Circularly polarized light is applied at constant fluency within a 3 $\mu$m long window around the domain wall. The blow-up shows the computed structure of the DW and schematically shows the spin up or spin down spin-polarized charge carriers generated by the circularly polarized light.}
\end{figure}

\begin{figure}[H]
\hspace*{-2 cm}\includegraphics[scale=0.7]{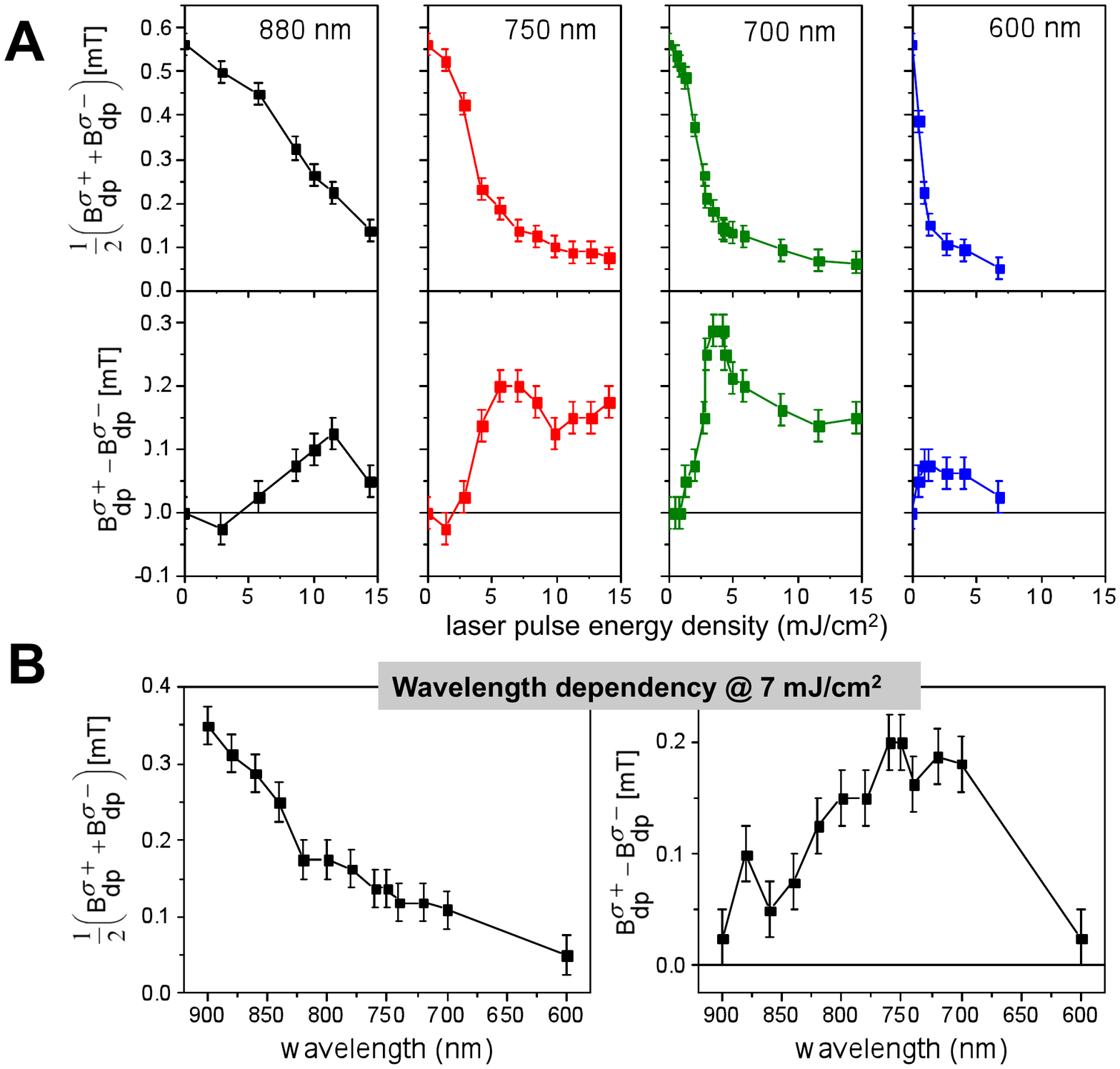}
\caption{
(A) Average and difference of the depinning field  $B_{dp}$  for $\sigma ^+$ and $\sigma ^-$ vs. LP energy density at various wavelengths at 90~K sample temperature. The average $ 1/2 [B_{dp}(\sigma ^+)+B_{dp}(\sigma ^-)] $ indicates the reduction of the geometrical pinning due to helicity independent LP heating. The difference $B_{dp}(\sigma ^+)-B_{dp}(\sigma ^-)$ shows the effect of the oSTT on the DW motion.
(B) The average at a fixed LP energy density of 7~mJ/cm$^2$ identifies the reduction of geometrical DW pinning with increasing LP heating due to enhanced absorption at higher photon energy.  On the other hand,  the difference $B_{dp}(\sigma ^+)-B_{dp}(\sigma ^-)$ shows that the oSTT efficiency is highest when the photon energy is close to the band gap of GaAs and it is strongly suppressed when photo-electrons are generated from the  spin-split-off band at high energy with $\lambda = 600$~nm.} 
\end{figure}

\begin{figure}[H]
\hspace*{0.5cm}\includegraphics[scale=0.6]{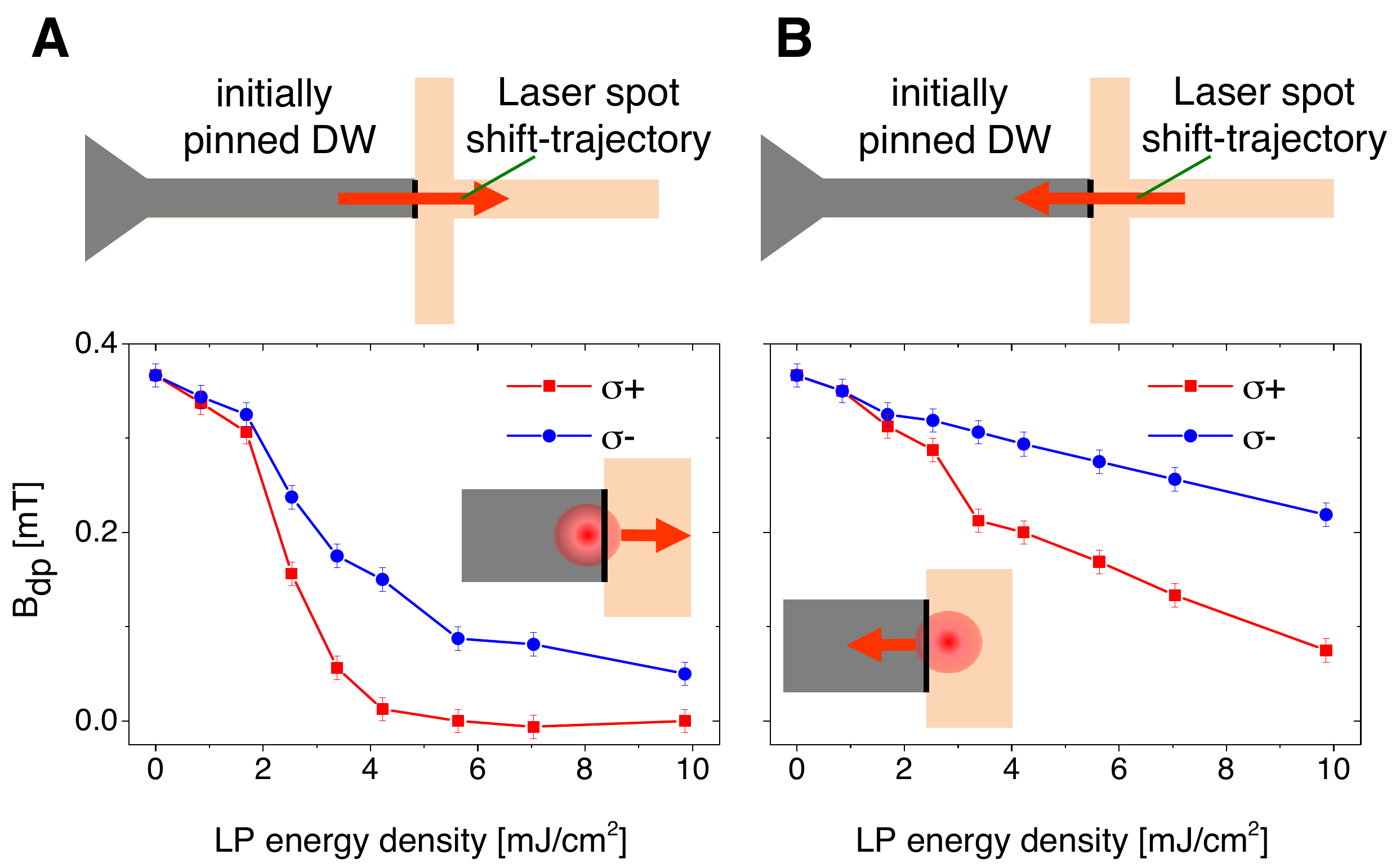}
\caption{Depinning fields $B_{dp}(\sigma ^+)$ and $B_{dp}(\sigma ^-)$ versus LP energy density, (A), in case that the focused LP spot approaches from the narrow bar, and (B), when the LP spot approaches from the wider cross area.} 
\end{figure}

\begin{figure}[H]
\hspace*{0.cm}\includegraphics[scale=0.7]{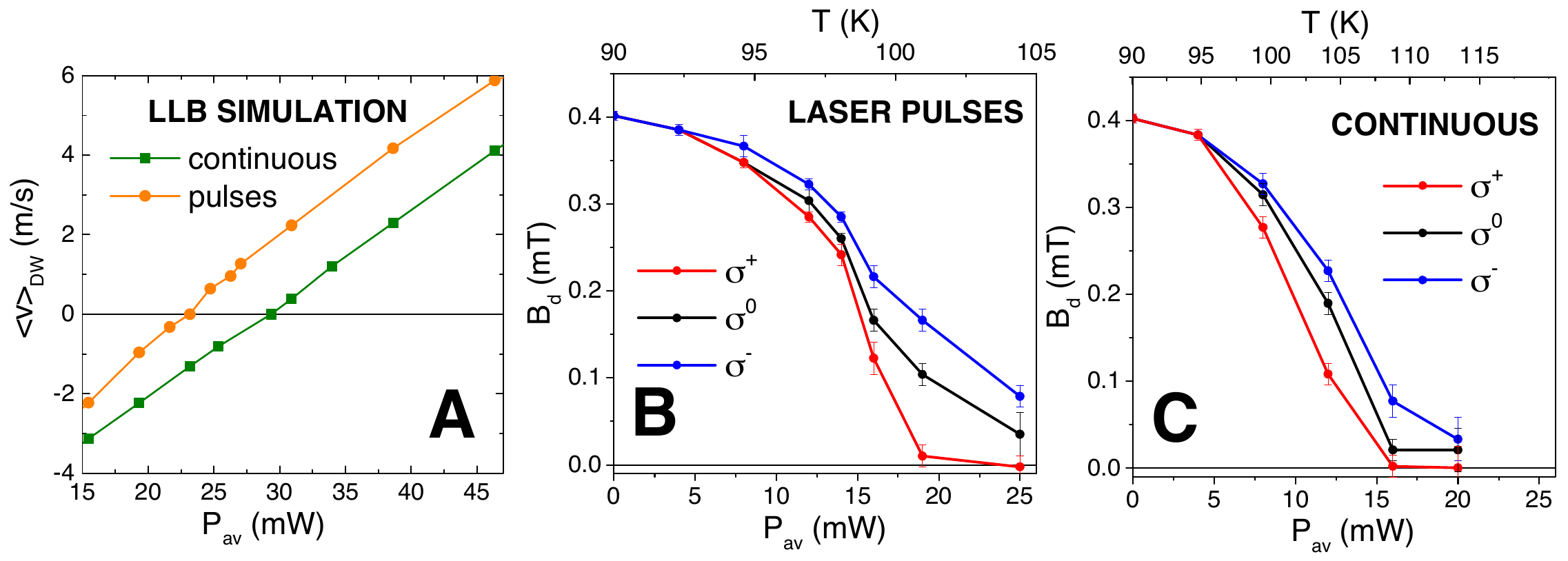}
\caption{(A) Averaged DW velocities vs. $P_{av}$ at the position of maximal restoring field for DW propagation driven by oSTT  with ultrashort LP excitation (orange) and cw-excition (green). The averaged velocity is deduced from simulations of DW propagation based on the LLB approach.
Depinning field $B_{dp}$ vs. $P_{av}$ for circularly ($\sigma ^+$),  ($\sigma ^-$) and  linearly polarized ($\sigma ^0$) laser light in case of LP- (B) and cw-(C) excitation. We have assigned an effective sample temperature deduced from the comparison between temperature dependent measurements of $B_{dp}$ without laser irradiation and measurements of $B_{dp}(\sigma ^0)$ vs. $P_{av}$  laser power and at fixed base temperature.} 
\end{figure}

\begin{figure}[H]
\hspace*{0.5cm}\includegraphics[scale=0.7]{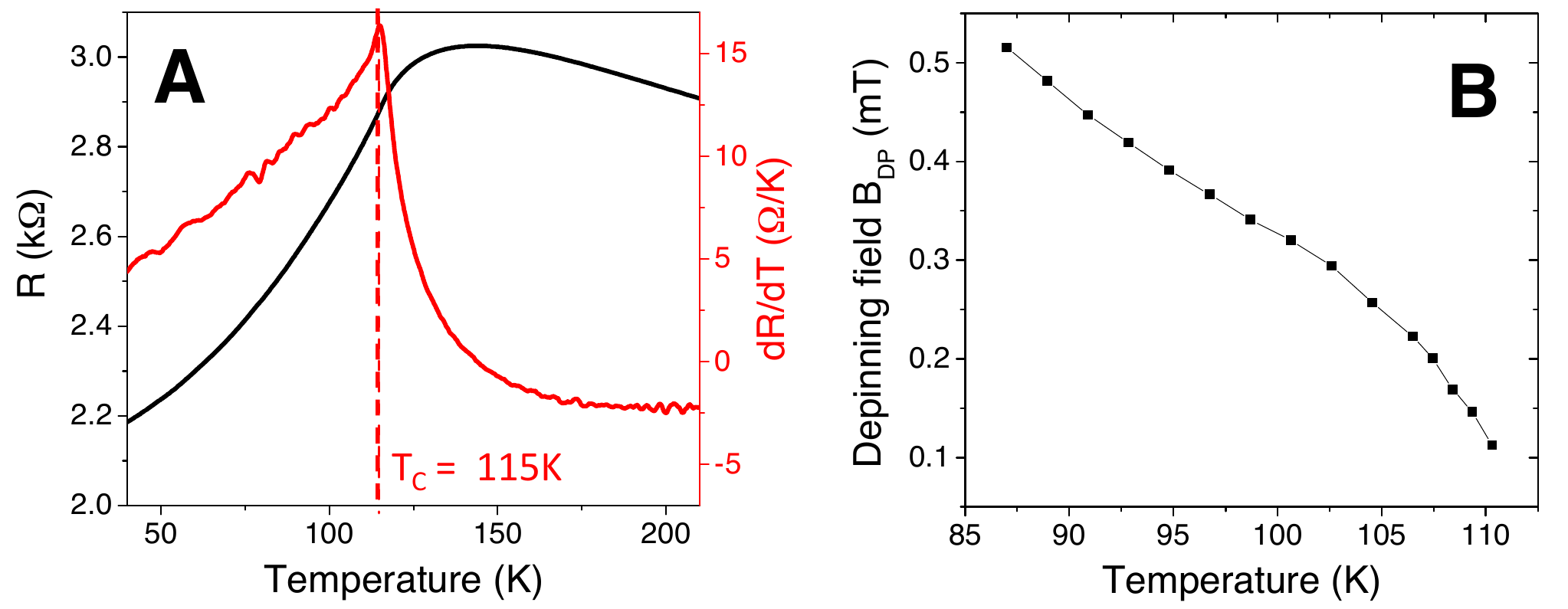}

\caption{(A) Temperature dependence of the resistance of the magnetic bar and $dR/dT$ identifying a Curie temperature of $Tc = 115$~K.
(B) Depinning field $B_{dp}$ without laser irradiation as a function of temperature.}
\end{figure}

\bibliographystyle{Science}
\bibliography{refs_JW,refs_TJ}

\begin{thebibliography}{10}
\bibitem{heating_experiment}
We estimate the LP heating related temperature increase by comparing $B_{DP}
  (\sigma^{0},$ LP energy density, $T=90K$) measured at constant $90K$ base
  temperature with the temperature dependence of $B_{DP}^{0}(T)$ without LP
  irradiation. We found that for LP energy densities of up to 35~mJ/cm$^2$, the
  temperature increase does not exceed the Curie temperature of $\sim 115~K$ of
  the GaMnAsP film.
\end{thebibliography}

\end{document}